\documentclass[12pt, multi, tikz]{spieman}  
\usepackage{amsmath,amsfonts,amssymb}
\usepackage{graphicx}
\usepackage{setspace}
\usepackage{tocloft}

\title{Neural network for estimation of optical characteristics of optically active and turbid scattering media}

\author[a]{Ali Alavi}
\affil[a]{School of Electrical and Computer engineering, College of Engineering, University of Tehran}

\cftpagenumbersoff{figure}
\cftpagenumbersoff{table} 
\begin{document} 
\maketitle

\begin{abstract}
One native source of quality deterioration in medical imaging, and especially in our case optical coherence tomography (OCT), is the turbid biological media in which photon does not take a predictable path and many scattering events would influence the effective path length and change the polarization of polarized light. This inherent problem would cause imaging errors even in the case of high resolution of interferometric methods. To address this problem and considering the inherent random nature of this problem, in the last decades some methods including Monte Carlo simulation for OCT was proposed. In this approach simulation would give us a one on one comparison of underlying physical structure and its OCT imaging counterpart. Although its goal was to give the practitioners a better understanding of underlying structure, it lacks in providing a comprehensive approach to increase the accuracy and imaging quality of OCT imaging and would only provide a set of examples on how imaging method might falter. To mitigate this problem and to demonstrate a new approach to improve the medical imaging without changing any hardware, we introduce a new pipeline consisting of Monte Carlo simulation followed by a deep neural network.
\end{abstract}


{\noindent \footnotesize\textbf{*}Ali Alavi,  \linkable{ali.alavi@ut.ac.ir} }

\begin{spacing}{2}   

\section{Introduction}
With the advent of deep learning methods in the last decade, biomedical imaging has seen special attention in the methods for segmentation and image quality enhancement.Although these methods could be beneficial in real world settings, it's quite difficult to estimate their accuracy and applicability, given the underlying physical structure is unknown.\\
To address this problem we propose an end to end approach to create a OCT results and its corresponding physical structure. Given these two pair we can train the network to infer the underlying structure, given the OCT imaging simulation results.\\
In section \ref{sect:pastworks} we will review the past works on simulating the OCT and the contribution of these methods and their drawbacks and their applicability. In section \ref{sect:montecarlosim} we will get into the details of Monte Carlo simulation of our work and different techniques we used and how we verified our results. In \ref{sect:neuralnetwork} we define our neural network and how it was trained. In \ref{sect:results} we present the results from Monte Carlo simulation and neural network results. In \ref{sect:conclusion} we give a summery of our work and give a glimpse of how it can be improved upon and be used for real world settings.

\section{Previous Works}
\label{sect:pastworks}
One of the first works on Monte Carlo simulation of OCT was in \cite{old1}. In this work the most straightforward implementation OCT simulation was presented. 
Another method used for simulation of light propagation for flurescence molecular tomography and bioluminescence tomography applications \cite{old10} is Radiative Transfer Equation (RTE) \cite{old11}. One of the main drawbacks of RTE method is very high computational load of this method. Some methods have been proposed to estimate the solution and reduce the computation cost, including diffusion approximation \cite{old12} and spherical harmonic method \cite{old15}. One of the main advantages of Monte Carlo method is the feasibility of simulation of turbid and complex media. The main drawbacks of Monte Carlo method are false scattering and uncertainty and error in the final result which can be mitigated by increasing the iterations.\\
To be able to simulate the result on mesh structures \cite{old16} and \cite{old18} have proposed voxel based methods.

In \cite{old39} the RTE method for light propagation in a simple biological media is investigated. In \cite{old38} the effect of Gaussian beam shape of LASER light in Monte Carlo simulation is discussed. Even though in high accuracy simulation, the Gaussian beam shape of source should be accounted for, because it's mainly important until the first scattering, its effect is minimal.\\
The effect of depolarization and change of polarization of light of polarized light incident unto the media is investigated in Polarization Sensitive Optical Coherence Tomography (PSOCT).

.
\section{Monte Carlo Simulation}
\label{sect:montecarlosim}
Before we get into the Monte Carlo simulation of polarized light we should consider the terms under which the Monte Carlo simulation is valid. To be able to use the transfer matrix, scattering matrix and linear propagation direction we should be able to use geometric optics. For this reason all the discontinuities and particles of interaction should be much less than the light wavelength. When this term is not satisfied, we should solve a scattering problem for the incident wave and calculate the scattering direction probability in this way. Also when the scattering from a single particle is being investigated, the effect of all other particles are neglected. To be able to use this simplification, the density of scattering particles in the media should be low.\\
Monte Carlo simulation of polarized light follows a straight forward convention. First we will discuss the general procedure of light propagation and scattering. Then we will investigate the changes of light polarization in turbid media.\\
In each step for each photon first the step length is determined according the Beer-Lambert's extinction law. The photon weight is adjusted at the end of each step according the covered distance and extinction factor.\\
To account for the change of polarization we follow the method described in \cite{rado}. In this paper the Coherent Back Scattering (CBS) formula is used to take the change of polarization into account. In this method the xx, xy, helicity preserving and diagonal polarization are calculated as \ref{eq:ppol}.

\begin{equation}
\label{eq:ppol}
\begin{split}
p_{xx}(x_s, y_s)=\sum_{n}^{}W[1+\frac{Q(x_s, y_s)}{I(x_s, y_s)}],\\
p_{xy}(x_s, y_s)=\sum_{n}^{}W[1-\frac{Q(x_s, y_s)}{I(x_s, y_s)}],\\
p_{++}(x_s, y_s)=\sum_{n}^{}W[1+\frac{V(x_s, y_s)}{I(x_s, y_s)}],\\
p_{+-}(x_s, y_s)=\sum_{n}^{}W[1-\frac{V(x_s, y_s)}{I(x_s, y_s)}]
\end{split}
\end{equation}
Due to scattering and optical activity, the degree of polarization will change through the simulation. The degree of polarization is defined through \ref{eq:doc}.

\begin{equation}
\label{eq:doc}
DOC=\frac{2Real[E_{\odot}(x_s, y_s) E_{\otimes}^*(x_s, y_s)]}{|E_{\odot}(x_s, y_s)|^2+| E_{\otimes}^*(x_s, y_s)|^2}
\end{equation}

Taking \ref{eq:doc} into \ref{eq:ppol} will yield the equation \ref{eq:orthodoc}.

\begin{equation}
\label{eq:orthodoc}
\begin{split}
p_{xy}(x_s, y_s).pc_{xy}(x_s, y_s)=\sum_{n}^{}W.DOC_{xy}[1-\frac{Q(x_s, y_s)}{I(x_s, y_s)}],\\
p_{+-}(x_s, y_s).pc_{+-}(x_s, y_s)=\sum_{n}^{}W.DOC_{+-}[1-\frac{V(x_s, y_s)}{I(x_s, y_s)}]
\end{split}
\end{equation}
Also we have to take the effects of scattering into account. After each scattering event the Jones matrix is multiplied by the scattering matrix \ref{eq:scatmat}.

\begin{equation}
\label{eq:scatmat}
\begin{bmatrix}
S_2(\theta, \phi) & S_3(\theta, \phi) \\
S_4(\theta, \phi) & S_1(\theta, \phi)
\end{bmatrix}
\end{equation}
In this work the scattering from biological particles are simplified to scattering from a symmetrical sphere. Scattering from this sphere is defined according to the Mie theory. The scattering probability distribution is defined according to \ref{eq:phasef}. 

\begin{equation}
\label{eq:phasef}
P(\theta, \phi)= \frac{S_{11}(\theta).I_o + S_{12}(\theta).(Q_ocos 2\phi + U_o sin 2\phi)}{\int_0^{2\pi} \int_0^{\pi} [S_{11}(\theta)I_o+S_{12}(\theta)(Q_o cos 2\phi + U_o sin 2\phi)]sin\theta d\theta d\phi}
\end{equation}
\begin{equation}
\label{eq:meul}
\begin{split}
S_{11}(\theta)=\frac{|S_2(\theta)|^2+|S_1(\theta)|^2}{2}\\
S_{12}(\theta)=\frac{|S_2(\theta)|^2-|S_1(\theta)|^2}{2}
\end{split}
\end{equation}

Another effect we should take into account is the birefringence and optical activity. In \cite{rado} these steps are discussed in details. In this part to take the effect of both phenomena the Jones N matrix \cite{rado34} formalism is used.\\
To speed up the simulation, partial photon \cite{rado} was used. Due to the large number of required simulations and taking the iid nature of each photons, in respect to each others, into account, it is evident we can use the parallel programming paradigm to speed up the simulation. There are two main approaches to parallel programming, multiprocessor programming on CPU and utilization of Nvidia's CUDA \cite{cuda} library for parallel programming on GPU.\\
To speed up the programming, Numba \cite{numba} library with backend of CUDA in python was used. By this approach the simulation time reduces by orders of magnitude.\\
To primarily verify the Monte Carlo simulation results, we first get the reflection by degree of a perpendicular incident light to a turbid media. This 
reflection can be calculated form \cite{bar} results. The reflection of TE wave is in equation \ref{eq:reflect}.

\begin{equation}
\label{eq:reflect}
r_{hs}^{TE}= \frac{\mu_{eff}^{TE}(\theta_i)k_z^i - k_z^{eff}}{\mu_{eff}^{TE}(\theta_i)k_z^i + k_z^{eff}}
\end{equation}

In which the effective quantities are calculated by solving the Green's dyadic equations and calculating the aggregate effect of half space consist of multitudes of planes with spherical scattering particles using Ewal-Oseen \cite{bar} formalism \ref{eq:effect}.

\begin{equation}
\label{eq:effect}
\begin{split}
\mu_{eff}^{TE}(\theta_i) = 1+i\gamma \frac{S_{-}^1(\theta_i)}{cos^2\theta_i} \\
\epsilon_{eff}^{TE}(\theta_i) = 1+i\gamma[2S_{+}^1(\theta_i) - S_{-}^1(\theta_i)tan^2(\theta_i)\\
S_{+}^m(\theta_i)= \frac{1}{2}[S(0) + S_m(\pi - 2\theta_i)]\\
S_-^m(\theta_i) = S(0) - S_m(\pi-2\theta_i)
\end{split}
\end{equation}

\begin{figure}
\centerline{\includegraphics[width=0.8\textwidth]{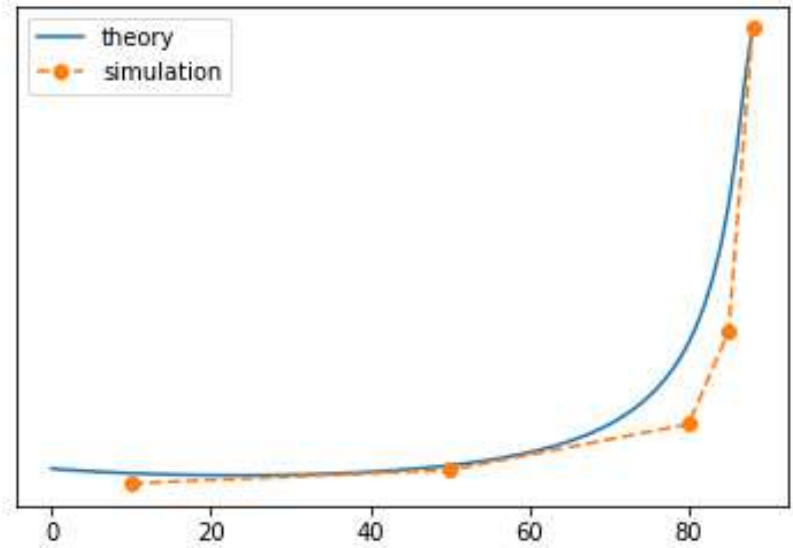}}
\caption{%
The reflection ration of incident light for different degree of reflections for perpendicular incident light.
}
\label{fig:reflecte}
\end{figure}

We compare this result with simulation output. In \ref{fig:reflecte} the comparison of theoretical and simulation results are shown.

\section{Neural Network}
\label{sect:neuralnetwork}

\begin{figure}
\begin{center}
\begin{tabular}{c}
\includegraphics[height=6cm]{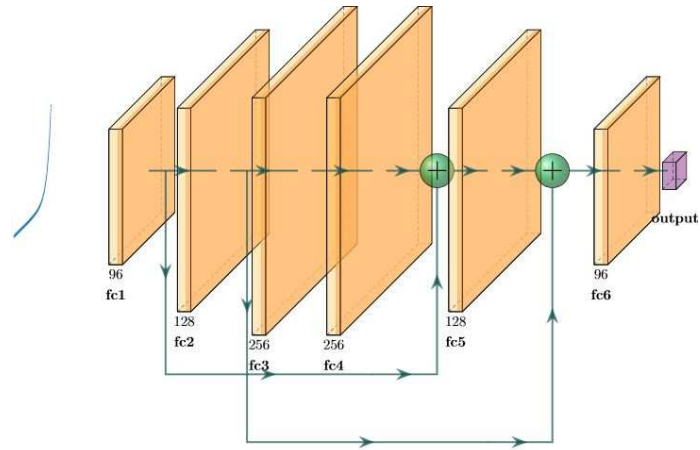}
\end{tabular}
\end{center}
\caption 
{ \label{fig:network}
The neural network consists of a variational encoder with residual connections architecture. The input to this network is the PSOCT measurements and the output is the optical properties.} 
\end{figure}

Given the Monte Carlo simulation results for polarization sensitive optical coherence tomography and having the underlying physical structure of the system under investigation, we know try to use a neural network to model the OCT imaging procedure and the physical system altogether.\\
What we do in this paper is infer the optical properties of a single layer turbid system with optically active, birefringent, scattering media from the PSOCT simulation measurements for different polarizations. The defined problem in this section is the inference from a higher dimensional measurement space to low dimension optical property space. For this purpose we propose an auto encoder with residual connections architecture. This network gives the ability to infer the optical properties from the latent space of measurements and residual connections to infer higher frequency data for capturing the information in higher frequency part of the data \ref{fig:network}.
The network was trained using ADAM optimizer and annealing learning rate after each 500 steps. The training and test split is 0.7 and 0.3 respectively. We use L2 loss of the error of in comparison to real optical property.

\section{Results}
\label{sect:results}
\begin{figure}
\begin{center}
\begin{tabular}{c}
\includegraphics[height=6cm]{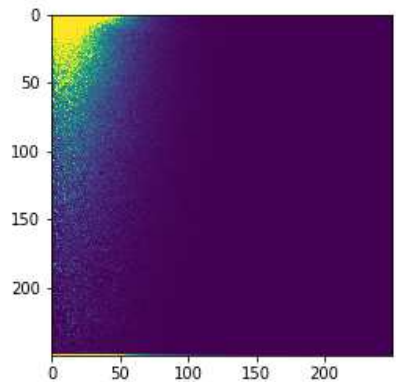}
\end{tabular}
\end{center}
\caption 
{ \label{fig:colin}
The simulation result for co polarization of linearly polarized incident light. Vertical axis photon depth and horizontal axis is different radius from incident position. The photon wavelength is 632 nm and each segment shows 4 micrometers.} 
\end{figure}

\begin{figure}
\begin{center}
\begin{tabular}{c}
\includegraphics[height=6cm]{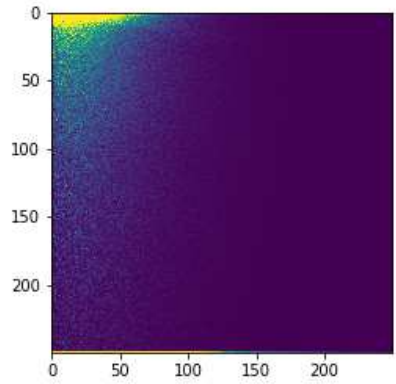}
\end{tabular}
\end{center}
\caption 
{ \label{fig:cocirc}
The simulation result for co polarization of circularly polarized incident light. Vertical axis photon depth and horizontal axis is different radius from incident position. The photon wavelength is 632 nm and each segment shows 4 micrometers.} 
\end{figure}

Monte Carlo simulation of a single layer media was done using the methods discussed in \ref{sect:montecarlosim}. The Mie scattering factors for sphere particles with 70 nm radius was calculated. Light with linear and circular polarization was incident to the media and in \ref{fig:colin} and \ref{fig:cocirc} the resulting photon which has reached the detector are shown.

These figures show the loss of co polarization due to scattering in the media.

\begin{figure}
\begin{center}
\begin{tabular}{c}
\includegraphics[height=6cm]{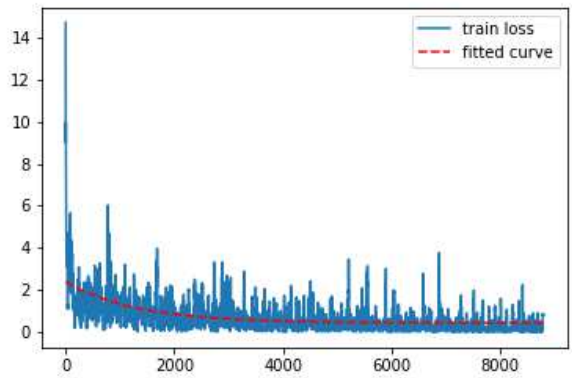}
\end{tabular}
\end{center}
\caption 
{ \label{fig:trainloss}
L2 train loss during training on the result of Monte Carlo simulation results.} 
\end{figure}

\begin{figure}
\begin{center}
\begin{tabular}{c}
\includegraphics[height=6cm]{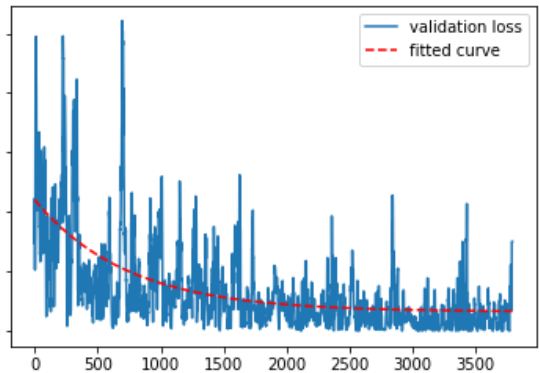}
\end{tabular}
\end{center}
\caption 
{ \label{fig:valloss}
L2 validation loss during training on the result of Monte Carlo simulation results.} 
\end{figure}

The figures \ref{fig:trainloss} and \ref{fig:valloss} show the training and validation loss during train steps. As it is evident, The network can predict the optical property, refractive index in this case, with good accuracy.
\section{Conclusion}
\label{sect:conclusion}
In this work we introduced a new approach to increase the accuracy and resolution of OCT methods without any hardware addition to the OCT setup. In this end to end approach the result of OCT imaging was simulated using Monte Carlo simulation on GPU and its results were used for neural network training. The network was trained on a simple physical media, but the method can be extended, given proper hardware is in hand. Adding the ray tracing and generating procedurally generated structures and training the network on this data, can enable the real world application and open a new venue for physically based resolution enhancement for biomedical imaging methods, especially OCT.


\bibliography{article}   

\begin{thebibliography}{10}

\bibitem{old1}
L.~Wang, S.~L. Jacquesa, and L.~Zheng, ``Ml - monte carlo modeling of light
  transport in multi-layered tissues,'' {\em Computer Methods and Programs in
  Biomedicine} {\bf 47}, 131--146  (1995).

\bibitem{old10}
V.~C, P.~andRazanskyD, PerrimonN, {\em et~al.}, ``In vivo imaging of drosophila
  melanogaster pupae with mesoscopic fluorescence tomography,'' {\em Nat.
  Methods} {\bf 5}, 45--7  (2008).

\bibitem{old11}
B.~I. W, {\em The Mathematics of Radiative Transfer}, Cambridge University
  Press  (1960).

\bibitem{old12}
A.~S, {\em Optical tomography in medical imaging Inverse Problems}, biomedical
  optics  (1999).

\bibitem{old15}
K.~A. D, N.~V, and H.~A. H, ``The inverse source problem based on the radiative
  transfer equation in optical molecular imaging,'' {\em J. Comput. Phys.} {\bf
  202}, 323–45  (2005).

\bibitem{old16}
P.~T. J, B.~J. K, C.~E. K, {\em et~al.}, ``A threedimensional modular adaptable
  grid numerical model for light propagation during laser irradiation of skin
  tissue,'' {\em IEEE J. Sel. Top. Quantum} {\bf 2}, 934–42  (1996).

\bibitem{old18}
H.~Shen and G.~Wang, ``A tetrahedron-based inhomogeneous monte carlo optical
  simulator,'' {\em Phys. Med. Biol.} {\bf 55}, 947–962  (2010).

\bibitem{old39}
J.~M. Schmitt and A.~Knuttel, ``Model of optical coherence tomography of
  heterogeneous tissue,'' {\em J. Opt. Soc. Am.} {\bf 14}, 1231–1242  (1997).

\bibitem{old38}
B.~H. Hokr, J.~N. Bixler, G.~Elpers, {\em et~al.}, ``Modeling focusing gaussian
  beams in a turbid medium with monte carlo simulations,'' {\em Optics Express}
  {\bf 23}(7)  (2015).

\bibitem{rado}
R.~AJ, R.~JD, C.~IR, {\em et~al.}, ``Open source software for electric field
  monte carlo simulation of coherent backscattering in biological media
  containing birefringence,'' {\em J Biomed Opt.} {\bf 17}(11)  (2012).

\bibitem{rado34}
R.~C. Jones, ``A new calculus for the treatment of optical systems. vii.
  properties of the n-matrices,'' {\em J. Opt. Soc. Am.} {\bf 38}(8), 671–683
   (1948).

\bibitem{cuda}
NVIDIA, P.~Vingelmann, and F.~H. Fitzek, ``Cuda, release: 10.2.89,''  (2020).

\bibitem{numba}
C.~R. Harris, K.~J. Millman, S.~J. van~der Walt, {\em et~al.}, ``Array
  programming with {NumPy},'' {\em Nature} {\bf 585}, 357--362  (2020).

\bibitem{bar}
R.~G. Barrera and A.~Garcia-Valenzuela, ``Coherent reflectance in a system of
  random mie scatterers and its relation to the effective-medium approach,''
  {\em J. Opt. Soc. Am. A} {\bf 20}, 296--311  (2003).

\end{thebibliography}
\bibliographystyle{spiejour}   


\vspace{1ex}
\noindent Biographies and photographs of the other authors are not available.

\listoffigures

\end{spacing}

\end{document}